\documentclass{Interspeech2024}

\usepackage{amsmath}
\usepackage{graphicx}
\usepackage{wrapfig}
\usepackage{xcolor}
\usepackage{multirow}
\usepackage{numprint}
\usepackage{booktabs}

\let\templatethebibliography=\thebibliography
\let\templateendthebibliography=\endthebibliography

\usepackage[numbers]{natbib}

\let\thebibliography=\templatethebibliography
\let\endthebibliography=\templateendthebibliography

\everypar{\looseness=-1}
\interspeechcameraready

\title{Transformer-based Model for ASR N-Best Rescoring and Rewriting}
\name{Iwen E. Kang, Christophe Van Gysel, Man-Hung Siu}

\address{Apple}
\email{\{ekang, cvangysel, manhung\_siu\}@apple.com}


\begin{document}

\maketitle

\begin{abstract}
Voice assistants increasingly use on-device Automatic Speech Recognition (ASR) to ensure speed and privacy. However, due to resource constraints on the device, queries pertaining to complex information domains often require further processing by a search engine. For such applications, we propose a novel Transformer based model capable of rescoring and rewriting, by exploring full context of the N-best hypotheses in parallel. We also propose a new discriminative sequence training objective that can work well for both rescore and rewrite tasks.
We show that our Rescore+Rewrite model outperforms the Rescore-only baseline, and achieves up to an average $8.6$\% relative Word Error Rate (WER) reduction over the ASR system by itself. 
\end{abstract}

\section{Introduction}

ASR converts user spoken audio into word sequences and often represents the word sequence as a ranked list of N hypotheses, known as the N-best list. The top hypothesis (1-best) is used as the recognized input for downstream tasks. To ensure speed and privacy, we can implement the ASR recognizer as an on-device component, keeping user-spoken audio within the device. On device ASR imposes constraints on model size that still works well for personal tasks like ``call mom'' or ``set alarm'', however, it can be sub-optimal for entity-rich queries in knowledge domains \citep{pusateri2019connecting,gondala2021error}, such as ``play Yesterday by Beatles''. Given these knowledge queries are eventually processed in the server \citep{vangysel2023modeling,zhang2024rescoring,sannigrahi2024llmgen}, ASR can also be improved downstream by either re-ranking the N-best hypotheses, i.e., ``rescoring'', or overriding the 1-best with its predicted corrections, i.e., ``rewriting''.

Conventional N-best rescoring often involves re-ranking based on a per hypothesis ``score'' computed individually \cite{Katz1987backoff,guo2019spelling,huang2019empirical,xu2022rescorebert,shin2019effective,fohr2021bert}. This approach typically interpolates with ASR acoustic scores for the second-pass rescoring. An interesting alternative is to explore the entire N-best list as input context for rescoring, and rewriting 1-best by leveraging joint N-best information. 

Using a transformer model for N-best rescoring has been proposed by others. \citet{guo2019spelling} proposed an LSTM-based Spell-Correction (SC) model to use N-best text data for error corrections.  
\citet{hrinchuk2020correction} proposed a vanilla Transformer model that operates similarly to neural machine translation (NMT) \cite{sutskever2014sequence}, specifically focused on rewriting without rescoring.
\citet{xu2022rescorebert} trained a BERT-based \cite{kenton2019bert} rescoring model with Minimum Word Error Rate (MWER) \cite{prabhavalkar2018minimum} loss. \citet{pandey2022lattention} proposed a rescoring model with attention to lattices. 
\citet{hu2021transformer,hu2022improving} proposed a Transformer-based rescoring model, which achieves better accuracy and latency than their previous LSTM-based counterpart \cite{sainath2019two}. Their ``Transformer Deliberation Rescorer'' model attends to both encoded audio and text hypotheses. 
\citet{Variani2020,Variani2022} proposed an N-best rescoring method using acoustic representations as inputs for optimizing ASR interpolation weights, and showed that Oracle Prediction, an edit-distance based adaptive weight optimization method, outperforms the non-adaptive weights.

Contrary to the related work above, which focused exclusively on either rescoring or rewriting task, in this paper, we propose a model capable of both rescoring and rewriting ASR hypotheses. This new model takes the full context of the N-best hypotheses in parallel.
Our proposed \textit{Transformer Rescore Attention} (TRA) model is illustrated in Figure~\ref{fig:transformer_rescore_attention}. We also propose a new discriminative sequence training objective, \textit{Matching Query Similarity Distribution} (MQSD), that can work well with cross-entropy based training to perform both rescore and rewrite tasks.
Our work is different from \cite{Variani2020,Variani2022}: a) Their model requires acoustic representations as inputs, while our model does NOT. Our TRA can operate as a standalone model outside of on-device ASR, the acoustic representations never leaves the device hence preserving privacy. b) Our MQSD performs effectively for both rescoring/rewriting tasks, whereas their Oracle Prediction is solely utilized to train a model for optimizing adaptive weight interpolations.

We show that: (1) our TRA model trained with MQSD works well for both rescore and rewrite tasks; (2) As a standalone model, our Rescore+Rewrite model outperforms the Rescore-only baseline model; (3) As an external LM model for ASR weights interpolation, our TRA model also outperforms the 4-gram LM \cite{Katz1987backoff}.

\section{Models}

In certain cases, due to privacy \cite{coman2021reconstruct} or design limitations, the on-device ASR acoustic embeddings may not be available to downstream tasks that operate in the cloud. For example, the on-device encoder state has a length proportional to the audio, and hence, may result in a payload that would incur too much latency to transmit over the network.
Hence, we choose to simplify the “Transformer Deliberation Rescorer” \cite{hu2021transformer} by removing acoustic components from the model, and re-implement it as our baseline ``N-best Transformer Rescorer'' (TR) model.

\subsection{Transformer Rescorer (TR) Model}
\label{sec:methodology:baseline}

Our baseline TR model is trained using a combined cross-entropy and MWER loss function. The MWER tries to minimize the expected number of word errors over the N-best hypotheses expressed in this equation:
\begin{equation*}
  L_{MWER}(x, y^*) =  \sum\limits_{y_i\in B\left(x,N\right)} p\left(y_{i}|x\right) \cdot \left(W\left(y_i,y^*\right) - E_{N}\right)
\end{equation*}
where $B(x,N) = \{y_{1},... ,y_{N}\}$ is the set of N-best hypotheses for the input query $x$, $W(y_i, y^{*})$ is the number of word errors in a hypothesis $y_i$ relative to the ground-truth sequence $y^{*}$, $p(y_{i}|x)$ is the normalized probability for hypothesis $y_i$ given input query $x$ such that $\sum_{y_i} p(y_i|x) = 1$, $E_N$ is the averaged word errors over the N-best hypotheses.
This MWER loss is added to the Transformer per-token cross-entropy loss $L_{ce}$ (see equation 3 in \cite{prabhavalkar2018minimum}) with an interpolation hyper-parameter $\alpha$ as the combined loss: $L=L_{MWER} + \alpha L_{ce}$.    

At inference time, the normalized probability for each N-best hypothesis is generated in teacher-forcing mode \cite{sainath2019two,hu2021transformer}, i.e., the Transformer's decoder is used to process each hypothesis (without beam-search) to generate its sequence loss. This loss can be converted into a probability using the $sigmoid$ function, and then re-normalized over the probability sum of the N-best hypotheses. 

\subsection{Transformer Rescore Attention (TRA) Model}

\begin{figure*}[th]
  \centering
  \includegraphics[width=14cm]{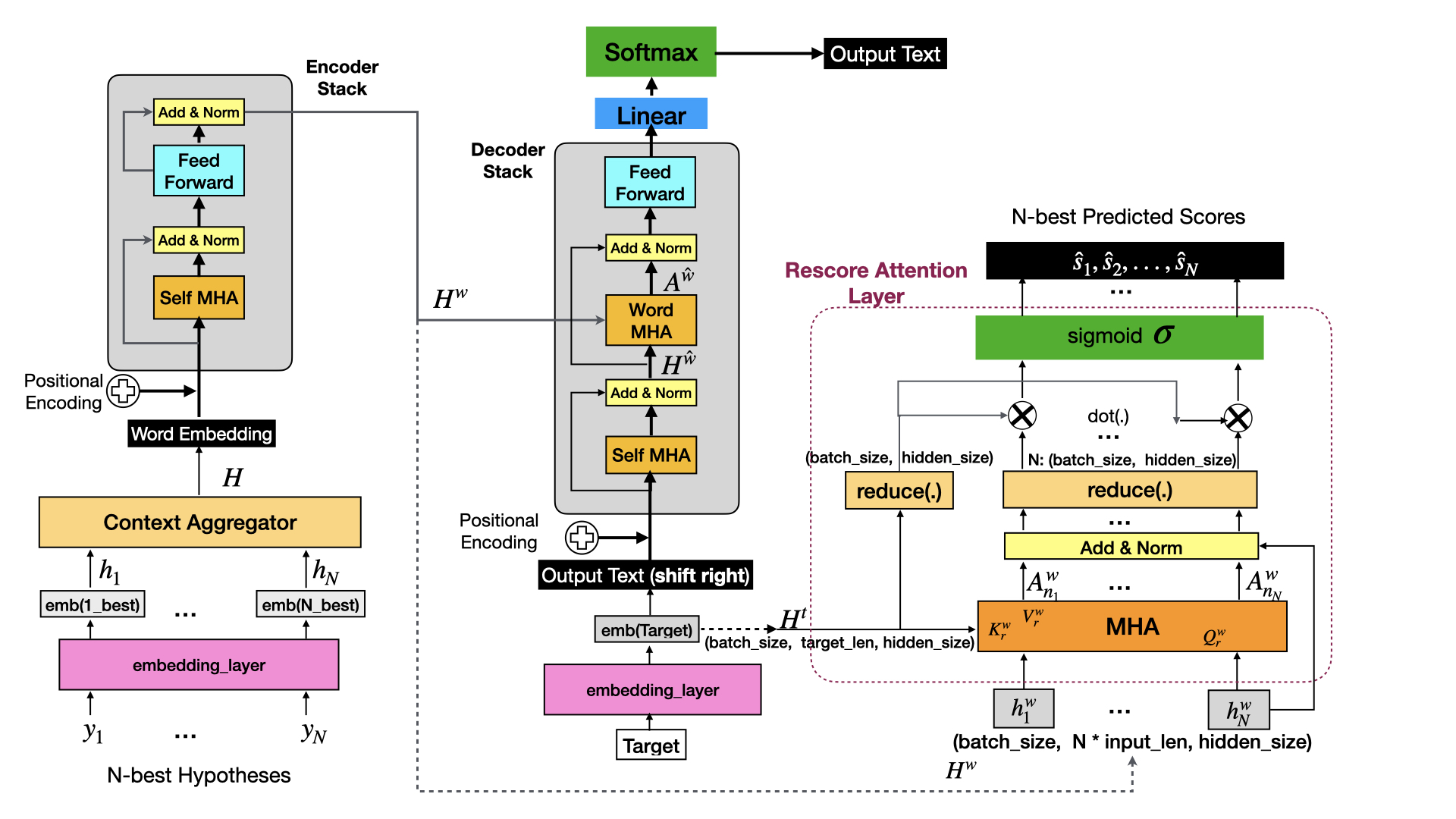}
  \caption{Transformer Rescore Attention (TRA) model. During training: the \textit{Target}, N-best and \textit{query similarity scores} are fed to TRA, the \textit{Target} sequence is shifted right as input for the decoder stack to compute per-token cross-entropy loss, the \textit{query similarity scores} are used as input against the N-best predicted scores to compute \textit{MQSD} loss. At inference time: the predicted ``Target'' is used to compute the N-best predicted scores for rescoring, the predicted output text can also be used to override the 1-best if its sequence loss exceeds a threshold.}
  \label{fig:transformer_rescore_attention}
\end{figure*}

We propose a \textit{Transformer Rescore Attention} (TRA) model by enhancing the TR baseline model with a \textit{Rescore-Attention Layer}, see Figure~\ref{fig:transformer_rescore_attention}.
The input of the model is a list of N-best hypotheses $\{y_{1},...,y_{N}\}$ from the ASR recognizer. Each hypothesis $y_i$ is tokenized into a sequence of subword tokens. The Transformer encoder and decoder stacks are similar to \cite{vaswani2017attention}. The \textit{embedding\_layer} maps each hypothesis $y_{i}$ to a high dimensional hidden embedding space $h_{i} \in R^{l \times d}$, where $l$ is the input sequence length and $d$ is the dimension of the hidden space.  
Each hypothesis from the same input query is padded to the same sequence length, and input sequences of the same N-best list size are grouped into the same batch during model training. 
The \textit{Context Aggregator} concatenates the N-best hypotheses along the sequence-length dimension, denoted as $\left[h_{1};...;h_{N} \right]$ with a concatenated sequence length $l_N = N * l$.
The concatenated N-best hidden vector $H = \left[ h_{1};...;h_{N} \right]$ is fed to the Transformer’s encoder stack to generate the N-best encoded vector $H^{w} = \left[ h^w_{1} ; ... ; h^w_{N} \right] \in R^{l_N \times d}$, which in turn is fed to the Transformer’s decoder stack to generate the cross-attention $A^{ \hat {w}}$ between the decoder’s self-attention output $H^{\hat{w}}$ and the N-best encoder output $H^{w}$.
Similar to \cite{vaswani2017attention}, this cross-attentions $A^{\hat {w}}$ is normalized and fed to a \textit{Feed-Forward} layer followed by a \textit{Linear} layer, then, a \textit{Softmax} layer to predict the next target token over a set of vocabulary, until it reaches a special end-of-sequence token \textit{eos}. We denote the predicted target sequence as $\hat {w} = \{\hat {w_1}, ..., \hat {w_t}\}$, and the target sequence length as $|\hat {w}|$. 

\subsection{Rescore Attention Layer}

This layer takes two inputs: the target embedding’s output $H^{t} \in R^{|\hat {w} | \times d}$ and the N-best encoded vector $H^{w}$. We first compute the N-best rescore cross-attention $A_{n}^{w}$ between $H^w$ and the target’s encoded output $H^t$ by:
\begin{equation*}
A_{n}^{w} = \text{Softmax}\left(\frac {\left(H^{w} Q_{r}^{w}\right)\left(H^{t} K_{r}^{w}\right)^{T}} {\sqrt{d}}\right) \left(H^{t} V_{r}^{w}\right)
\end{equation*}
where $Q_{r}^w$, $K_{r}^w$, $V_{r}^w \in R^{d \times d}$ are the parameters of the \textit{Rescore-Attention Layer}.
This N-best rescore cross-attention $A_{n}^{w}$ $\in R^{l_N \times d}$ provides a high-dimensional similarity measurement for each N-best hypothesis against the target sequence in different embedding subspaces via Multi-Head Attentions (MHA) \cite{vaswani2017attention}. %
This N-best rescore cross-attention is a concatenation of N attention blocks, i.e., $A_{n}^{w} = \left[ A^w_{n_1}; ... ;A^w_{n_N} \right] \in R^{l_N \times d}$, indexed by $i = 1, ..., N$, where each attention block $A_{n_i}^{w} \in R^{l \times d}$ is the $i$-th partition of of $A_{n}^{w}$ in the sequence-length dimension with length $l$.
The output of this MHA layer is normalized and then fed to a reduce sum function \textit{reduce($\cdot$)}, reduce by summing along the sequence-length dimension $l$, denoted by $\sum_l (X_{l, \cdot} \in R^{l \times \cdot})$. The same reduce function is applied to the target sequence. Finally, a scalar $\hat s_{i}$ for each N-best hypothesis $\{\hat s_{1}, ..., \hat s_{N}\}$ can be computed using $dot(\cdot)$ products and \textit{sigmoid} function $\sigma$ by: %
\begin{equation*}
\hat s_{i} = \sigma\left( \sum_l (H^t)_{l, \cdot} \cdot \sum_l \left(A^{w}_{n_i}\right)_{l, \cdot} \right).
\end{equation*}

\subsection{Matching Query Similarity Distribution (MQSD) Loss}

We propose an alternative to the MWER loss function that can work well with cross-entropy (CE) based training for models to perform both rescore and rewrite tasks. 
The \textit{Matching Query Similarity Distribution} (MQSD) loss is the cross-entropy loss of predicted scores over N-best hypotheses, denoted by $L_{MQSD}(x, y^*)$ as: %
\begin{equation*}
  -\sum\limits_{y_i\in B(x,N)} \left(\text{Softmax}\left(s_{i}\right) \cdot \log\left(\text{Softmax}\left(\hat s_{i}\right)\right)\right)
\end{equation*}
where $B(x,N) = \{y_{1},... ,y_{N}\}$ is the set of N-best hypotheses for the input query $x$; $s_{i} = (1 - wer(y_{i}, y^*))^{2}$ is the \textit{query similarity score} for hypothesis $y_{i}$, $wer(y_{i}, y^*)$ is the word error rate of hypothesis $y_i$ relative to the ground-truth sequence $y^*$, capped at 1.0; $\hat s_{i}$ is the predicted query similarity score for hypothesis $y_{i}$.
The \textit{query similarity score} is a similarity indicator between an N-best hypothesis and the target query. The score is in the range [0, 1], where higher score means higher similarity in edit distance and lower word error rate. 
Our MQSD loss is different from MWER: The MWER training minimizes the expected word errors over the N-best hypotheses through normalized N-best probabilities, while the goal of MQSD loss is to mimic the N-best \textit{query similarity scores} distribution in the ground-truth through predicted scores. 

We train TRA model with a combined objective: minimizing the Transformer’s cross-entropy loss $L_{ce}$ for the target token sequence and the cross-entropy loss $L_{MQSD}$ for the N-best scores, with a hyper-parameter $\lambda$ for interpolation: $L=L_{MQSD} + \lambda L_{ce}$.

\section{Experimental Setup}

\subsection{ASR system}
\label{section:experimental:system}

Our on-device ASR system uses a word-piece Conformer following \citep{Yao2021wenet} with wordpiece ouptuts and an external word-based LM trained with a large text corpus. Our decoder uses a similar strategy to \citep{Miao2015eesen} with an additional rescoring step \citep{Dolfing2001incremental,Lei2023acoustic}. All of our experiments operate on top of the N-best (with $N \leq 10$) list generated by the ASR system.

\subsection{Training and evaluation data}

\subsubsection{Training data}
\label{section:experimental:training}

We train domain expert models by mixing a large (95\%) \textit{synthetic} in-domain (music) training set with a small (5\%) all-domain \textit{annotated} training set. The 1.8M annotated queries are sampled anonymously from opted-in Voice Assistant (VA) queries across all domains and consist of text/audio pairs.
The 36M synthetic queries are obtained by enumerating in-domain entity data feeds with query templates, similar to \cite{VanGysel2022phirtn}.
While the 4-gram LM (\S\ref{section:experimental:methods:ngram}) is trained directly on the query texts, the Transformer models (\S\ref{section:experimental:methods:transformers}) require ASR N-bests lists for training. We obtain N-best lists by decoding audio using our ASR system (\S\ref{section:experimental:system}). For the 36M synthetic queries, we generate audio using Text-to-Speech (TTS).

\subsubsection{Evaluation sets}
\label{section:experimental:evaluation}

\begin{table}
\centering%
\renewcommand{\arraystretch}{0.75}%
% Caption should be above tables.
\caption{Overview of the evaluation sets.\label{tab:evaluation_sets}}
\begin{tabular}{llrr}
\toprule
\textbf{Collection} & \textbf{Sub-set} & \textbf{\# queries} & \textbf{avg. query length} \\
\midrule
\multirow{2}{*}{VA-2022} & All & \numprint{8,035} & 5.8 \\
& Music & \numprint{975} & 5.1 \\
\midrule
\multirow{2}{*}{VA-2023} & All & \numprint{11,998} & 6.0 \\
& Music & \numprint{1,100} & 5.9 \\
\bottomrule
\end{tabular}
\end{table}

We evaluate our models on randomly sampled, representative, and anonymized VA queries sampled across two years (2022 and 2023) where we compare across the entire population and the sub-population corresponding to music queries. %
Table~\ref{tab:evaluation_sets} shows an overview of our evaluation sets.

\subsection{Rescoring/rewriting methods under comparison}
\label{section:experimental:methods}

\subsubsection{Transformers}
\label{section:experimental:methods:transformers}

We use a 16k vocabulary SentencePiece (SP) \cite{kudo2018sentencepiece} to tokenize text input into a sequence of subword tokens. %
The SP model was trained on all queries in our training set (\S\ref{section:experimental:training}). %
Both TR and TRA models have 4 layers in the encoder stack, and 1 layer in the decoding stack, with MHA with 8 heads, hidden layers dimension 512, and 2048 units in the feed-forward layer for a total of 25M model parameters. The TRA model has an additional 1M parameters in the \textit{Rescore Attention Layer} which includes an 8-headed MHA.

\noindent %
\textbf{Training schedule.} %
Both TR and TRA models are trained up to 300,000 steps on an 8-GPUs machine, with batch size as large as possible (up to 30,000 tokens) per replica. We adopt an early-stopping criteria on a development-set (dev-set) to avoid over-fitting, the dev-set is a small split (about 1\%) from the annotated training set, and is evaluated every 5,000 steps. Similar to \cite{vaswani2017attention}, we use the same parameters for Adam optimizer and custom learning rate schedule, except with a lager value of 8000 for \textit{warmup\_steps}. We also use the same dropout ratio 0.1 for both Attention and ReLu layer. 
The baseline TR model is trained using the combined MWER objective $L = L_{MWER} + \alpha L_{ce}$,
following \cite{hu2021transformer} we set $\alpha = 0.01$ in our experiments.
Similarly, our TRA model is trained using the combined MQSD objective:
$L = L_{MQSD} + \lambda L_{ce}$. We set $\lambda =0.01$ in our experiments.

\noindent %
\textbf{Hyper-parameters.} %
\newcommand{\ThresholdR}{\text{threshold}_R}
\newcommand{\ThresholdW}{\text{threshold}_W}
There are 2 tunable parameters in our TRA model: %
1. \textit{$threshold_R$}, which triggers N-best rescoring only when the model's confidence score (log-probability) surpasses this threshold; %
2. \textit{$threshold_W$}, which dictates the conditions under which to rewrite the 1-best with the model's predicted text, applied when its sequence loss exceeds this threshold.
The thresholds are optimized by performing a grid-search over a range of possible values on a held-out dev-set, such that we have the lowest WER on in-domain dev-set while WER on the all-domain dev-set is not degraded. We searched for the best \textit{$threshold_R$} first and then constraint the search for \textit{$threshold_W$} $>$ \textit{$threshold_R$}.
Following this approach we set $\ThresholdR{}=-1.0$ and $\ThresholdW{}=-0.5$ in our TRA model. Same thresholds are applied to all TRA experiments. 
To avoid overriding 1-best when the N-best context is insufficient, we do not perform rewrite when $N=1$.
In our experiments, we present evaluation results for our TRA model in two configurations: with rewriting enabled (TRA-RW, rescore-rewrite) and without rewriting (TRA-R, rescore-only). 

\subsubsection{4-gram LM with Katz back-off}
\label{section:experimental:methods:ngram}

In addition to the Transformer models (\S\ref{section:experimental:methods:transformers}) trained on entire N-best lists, we also train a 4-gram back-off LM \citep{Katz1987backoff} on the reference text in our training set (\S\ref{section:experimental:training}). 3- and 4-grams with a count less than 2 are discarded. Good-Turing discounting is applied to 2-, 3- and 4-grams that have a frequency of 7 or less.
The 4-gram LM is combined with signals extracted from the ASR decoding process using linear interpolation to score the candidates in the N-best list, as described in the next section (\S\ref{section:experimental:interpolation}).
The advantage of the 4-gram LM is that it allows for estimation without needing to synthesize audio or run speech recognition to generate N-best lists, unlike what is required for Transformer models in \S\ref{section:experimental:methods:transformers}. However, the downside of this approach is that the 4-gram LM cannot utilize N-best list context to correct the unique error distribution of the ASR system.

\subsection{Interpolation with ASR decoding signals}
\label{section:experimental:interpolation}

In our second experiment set, we will combine the per-hypothesis signal from each of the models described above with the scores assigned by the ASR system (\S\ref{section:experimental:system}): the log-likelihood for the hypothesis given the input audio provided by the Conformer, and the log-likelihood of the hypothesis text under the on-device external LM \citep[Ch.~16]{Jurafsky2023slp3draft}.

For each of the rescore-only models above (TR, TRA-R, 4-gram LM), we include their log-probability as an additional signal. However, TRA-RW may generate a new hypothesis not part of the N-best list, and lacking corresponding ASR scores. In that case, the ASR-provided scores are assumed to be zero, and we introduce a fourth signal, \textit{lmCost+}, equal to the Transformer's generative log-likelihood.
Following \citet{zhang2024rescoring}, the various signals are combined through a linear combination, and we learn the weights via Powell's method \cite{Powell1964} against a separate development set sampled independently from the same population and with similar characteristics as the VA-2023 evaluation set. 

\section{Results}

\subsection{TR \& TRA}

\begin{table}[t]
\centering%
\renewcommand{\arraystretch}{0.75}%
% Caption should be above tables.
\caption{WER evaluation (\S\ref{section:experimental:evaluation}) of the Transformer models (\S\ref{section:experimental:methods:transformers}) with ASR N-best input. The average column depicts the arithmetic mean across the evaluation sets.\label{tab:results:standalone}}
\begin{tabular}{lrrrrrr}
\toprule
& \multicolumn{2}{c}{\textbf{VA-2022}} & \multicolumn{2}{c}{\textbf{VA-2023}} & \multicolumn{2}{c}{\textbf{Average}} \\
\textbf{Method} & \textbf{All} & \textbf{Music} & \textbf{All} & \textbf{Music} & \textbf{All} & \textbf{Music} \\
\midrule
ASR & 3.57 & 4.70 & 5.78 & 5.52 & 4.68 & 5.11 \\
\midrule
TR & 3.61 & 4.50 & 5.78 & \textbf{5.18} & 4.70 & 4.84 \\
TRA-R & 3.52 & 4.28 & \textbf{5.72} & 5.29 & \textbf{4.62} & 4.79 \\
TRA-RW & \textbf{3.51} & \textbf{3.98} & \textbf{5.72} & 5.36 & \textbf{4.62} & \textbf{4.67} \\
\bottomrule
\end{tabular}

\end{table}

Table~\ref{tab:results:standalone} shows the results of the Transformer models operating directly on the ASR N-best list texts. While TR and TRA-R perform rescoring of the ASR N-best list only, TRA-RW also has the ability to overwrite the 1-best. We find that TRA (both R and RW) perform best on the entire query population. However, for the music sub-population, results are more inconsistent and while TRA-RW performs best on VA-2022 (music), the TR model without the rescoring attention layer seems to perform better on the music subset of VA-2023 (closely followed by TRA-R). We suspect that this may be due to a drift in music entities popularity or user behavior, since the training data was generated in 2022, and hence, the TRA-RW model may be over-correcting to entities that are no longer as relevant in 2023. However, if we look at the average across both test sets, we see that both TRA-R/RW perform well on the entire query population, and TRA-RW provides a significant improvement ($8.6\%$ rel.) on the music sub-collection. This is most likely due to good coverage of the in-domain music synthetic queries (95\%) in our training set.

\subsection{Interpolation with ASR signals}

\begin{table}[t]
\centering%
\renewcommand{\arraystretch}{0.75}%
% Caption should be above tables.
\caption{WER on VA-2023 (\S\ref{section:experimental:evaluation}) by combining ASR decoding signals with signals from the rescoring/rewriting methods (\S\ref{section:experimental:methods}) using a linear model (\S\ref{section:experimental:interpolation}), where +$W^*$ denotes models with optimized weights. The relative improvement in WER over the ASR system is listed between brackets.\label{tab:results:interpolation}}
\begin{tabular}{lrr} 
\toprule
& \multicolumn{2}{c}{\textbf{VA-2023}} \\
\textbf{Method} & \multicolumn{1}{c}{\textbf{All}} & \multicolumn{1}{c}{\textbf{Music}} \\
\midrule
ASR & 5.78 \phantom{\footnotesize $(0.00\%)$} & 5.52 \phantom{\footnotesize $(0.00\%)$} \\
\midrule
4-gram LM +$W^*_{lm4}$  & 5.57 {\footnotesize $(3.63\%)$} & 5.26 {\footnotesize $(4.71\%)$} \\
TR +$W^*_{tr}$  & 5.57 {\footnotesize $(3.63\%)$} & 5.32 {\footnotesize $(3.62\%)$} \\
TRA-R +$W^*_{tra}$  & 5.47 {\footnotesize $(5.36\%)$} & \textbf{5.15 {\footnotesize $(6.70\%)$}} \\
\midrule
TRA-RW +$W^*_{tra+}$  & \textbf{5.46 {\footnotesize $(5.53\%)$}} & 5.23 {\footnotesize $(5.25\%)$} \\
\bottomrule
\end{tabular}
\end{table}

Table~\ref{tab:results:interpolation} shows the results when interpolating the log-probability provided by the various models with per-hypothesis signals extracted from the ASR decoding process (\S\ref{section:experimental:interpolation}). The incorporation of an additional external LM (\S\ref{section:experimental:methods:ngram}), which constitutes a traditional rescoring approach, yields an additional improvement of ${\sim}4\%$. However, compared to the 4-gram LM, the TRA models perform better and yield an relative improvement between $5.3\%$ and $6.7\%$ on the full collection, and the music sub-collection, resp.
This leads us to the following conclusion: while training a LM independently of the ASR system's error distribution is computationally simpler, as it eliminates the need for TTS and ASR decoding to generate N-best lists. However, training a Transformer using the ASR system's N-best list yields superior improvements because the Transformer approach can %
(a) make use of additional contextual information available in the N-best list (e.g., the number of hypotheses, segments that differ across hypotheses), and
(b) learn to correct error patterns made by the specific ASR system it was trained against.

\section{Conclusions}

In this paper, we proposed a novel Transformer-based model capable of rescoring and rewriting ASR hypotheses. We also propose a new discriminative sequence training objective MQSD that can work well with cross-entropy based training for both rescore and rewrite tasks. Given an N-best list as text input, our TRA model outputs both predicted text and \textit{query similarity scores} for N-best re-ranking. The predicted text can be used to override the top hypothesis if its sequence loss exceeds a threshold. 
As a standalone model, our TRA achieves up to $8.6\%$ WER improvement over the ASR baseline on in-domain test sets.
As an external LM for ASR interpolations, our TRA model also outperforms the 4-gram LM and the baseline TR model.
\\
\textbf{Acknowledgments.} %
We thank %
Russ Webb, %
Tatiana Likhomanenko, %
Tim Ng, %
Thiago Fraga-Silva, %
and the anonymous reviewers for their comments and feedback.

\bibliographystyle{IEEEtranN}
\bibliography{interspeech2024-rescore_rewrite}

\end{document}